\documentclass[aps,superbib,12pt,preprint]{revtex4}
\usepackage{amsfonts}
\usepackage{amsmath}
\usepackage{amssymb}
\usepackage{graphicx}

\begin{document}
\title{Spatial Interference: From Coherent To Incoherent }
\author{Su-Heng Zhang$^1$, Lu Gao$^1$, Jun Xiong$^1$, Li-Juan Feng$^1$, De-Zhong Cao$%
^2$, and Kaige Wang$^1\footnote{%
Corresponding author: wangkg@bnu.edu.cn}$}
\address{1.Department of Physics, Applied Optics Beijing Area Major Laboratory,\\
Beijing Normal University, Beijing 100875, China\\
2.Department of Physics, Yantai University, Yantai 264005, China}

\begin{abstract}
It is well known that direct observation of interference and diffraction
pattern in the intensity distribution requires a spatially coherent source.
Optical waves emitted from portions beyond the coherence area possess
statistically independent phases, and will degrade the interference pattern.
In this paper we show an optical interference experiment, which seems
contrary to our common knowledge, that the formation of the interference
pattern is related to a spatially incoherent light source. Our experimental
scheme is very similar to Gabor's original proposal of holography\cite
{gabor1}, just with an incoherent source replacing the coherent one. In the
statistical ensemble of the incoherent source, each sample field produces a
sample interference pattern between object wave and reference wave. These
patterns completely differ from each other due to the fluctuation of the
source field distribution. Surprisingly, the sum of a great number of sample
patterns exhibits explicitly an interference pattern, which contains all the
information of the object and is equivalent to a hologram in the coherent
light case. In this sense our approach would be valuable in holography and
other interference techniques for the case where coherent source is
unavailable, such as x-ray and electron sources.

\end{abstract}
\maketitle

At the early time when coherent sources were unavailable, interference
experiments were carried out by a thermal light source with the help of a
pinhole aperture. Though it can improve spatial coherence of the source, the
pinhole aperture, as a cost, eventually reduces the power of the source and
thus restricts the potential application of optical interferometric
techniques such as holography. The effort to realize interference with
chaotic light has been developed since the landmark experiment reported by
Hanbury-Brown and Twiss (HBT)\cite{hbt}. They realized that light from
different, completely uncorrelated portions of the star gives rise to an
interference effect which is visible in intensity correlations but not in
the intensities themselves, and proposed an intensity interferometer to
measure the angular size of distant stars. The intensity correlation
property of spatially incoherent light achieves significant development
recently in ghost interference and subwavelength interference\cite
{bd,zhu,wkg,shih,gigi1,wkg1}. The physics behind these effects is that each
point of a spatially incoherent source produces coherence of the field at
two separate positions, after having travelled different paths, and the
coherent information can be acquired through the intensity correlation
measurement of the two positions. Moreover, Ref. \cite{bo} reported that
phase and amplitude of the field correlation function of two positions can
be retrieved by a modified Young interferometer, instead of intensity
correlation measurement. There is still a challenging question whether, by
using an incoherent light source, the coherent information can be recorded
through intensity distribution itself?

When Dennis Gabor accomplished the first holography experiment, he did not
realized the fact that the requirment of spatial coherence can be avoided so
long as his interferometric scheme is somewhat modified. In this paper, we
propose such an interferometric scenario which is capable of carrying out
interference and diffraction in intensity observation using a spatially
incoherent source. The experimental setups of interferometer are sketched in
Fig. 1, which is similar to Gabor's original proposal of holography\cite
{gabor1}. The source field is divided into two sets: one illuminates an
object, called object wave, and the other acts as a reference wave. The
interference occurs at the outgoing beamsplitter BS$_2$ and can be recorded
by either one of two CCD cameras. The interference parts at the two outgoing
ports have a phase shift $\pi $ due to the reflection of the field. In order
to demonstrate primary principle simply, the object in the experiments is a
double-slit of slit width $b=125\mu m$ and spacing $d=310\mu m$. As a
proof-of-principle experiment, we first use a pseudo-thermal light source,
which is formed by passing a He-Ne laser beam of wavelength 632.8 $nm$
through a slowly rotating ground glass disk G. A step-motor moves the ground
glass each 80$ms$ in which CCD camera can register a frame of interference
pattern. The pattern fluctuates randomly by moving the ground glass. We
first consider the scheme of Fig. 1(a) in which two waves travel different
distances: $z_o=16cm$ for the object wave and $z_r=27cm$ for the reference
wave, and $|z_o-z_r|/c$ is less than the coherent time of the laser beam.
Experimental results of two-dimensional (2D) intensity patterns detected by
CCD$_1$ are summarized in Fig. 2. We can see that two single frames in Figs.
2(a) and 2(b) show irregular patterns. With the increasing of number of
frames to be averaged in Figs. 2(c)-2(g), the well-defined interference
pattern has emerged gradually.

The above experimental results can be readily explained by the fundamental
optics theory. Let $E_o(x)$ and $E_r(x)$ be the field distributions of the
object wave and the reference wave at the recording plane, respectively. The
interference term is given by $E_r^{*}(x)E_o(x)=\alpha _r^{*}\alpha _o\int
h_r^{*}(x,x_0)T(x_0^{\prime })h_o(x,x_0^{\prime
})E_s^{*}(x_0)E_s(x_0^{\prime })dx_0dx_0^{\prime }$, where $E_s(x_0)$ is the
source field at beamsplitter BS$_1$; $h_j(x,x_0)$ and $\alpha _j$ are the
impulse response functions between $E_s(x_0)$ and $E_j(x)$ ($j=o,r$) and the
attenuation constant in each path, respectively; $x_0$ and $x$ are the
transverse positions across the beam. A transmittance object $T(x)$ is
located close to BS$_1$ in the object path of the interferometer. For a
coherent source which wavefront $E_s(x_0)$ is stationary, the intensity
pattern $I(x)=|E_r(x)|^2+|E_o(x)|^2+[E_r^{*}(x)E_o(x)+$c.c.$]$ is stable.
However, if the source is a spatially incoherent field in which both the
amplitude and phase distributions fluctuate randomly, the interference
pattern $I(x)$ will fluctuate, too. This has been shown by the single frame
in Figs. 2(a) and 2(b).

The incoherent source field $E_s(x)$ is assumed to be quasi-monochromatic
and satisfies completely spatial incoherence $\langle
E_s^{*}(x)E_s(x^{\prime })\rangle =I_s\delta (x-x^{\prime })$. The
interference term in the statistical average can be obtained as
\begin{equation}
\left\langle E_r^{*}(x)E_o(x)\right\rangle =\alpha _r^{*}\alpha _oI_s\int
T(x_0)h_r^{*}(x,x_0)h_o(x,x_0)dx_0.  \label{2}
\end{equation}
The integration manifests that all portions of the source globally
contribute to the interference term. If both the object and reference waves
travel in exactly same configuration ($h_r=h_o$) as, for example, in an
usual interferometer, one immediately obtains a homogeneous distribution of
Eq. (\ref{2}). This used to be understood as an incoherent superposition
effect which washes out the information of the object. We now modify the
interferometer in an unbalanced way as shown in Fig. 1(a). For the moment we
assume that the source beam has temporal coherence. Hence Eq.(\ref{2}) is
still valid under such an appropriate path difference that $\langle
E_s^{*}(x,t)E_s(x^{\prime },t-|z_o-z_r|/c)\rangle \approx $ $\langle
E_s^{*}(x)E_s(x^{\prime })\rangle $. In the paraxial propagation, the
impulse response function for a free path $z_j$ is given by $h_j(x,x_0)=%
\sqrt{k/(i2\pi z_j)}\exp \left[ ikz_j+ik(x-x_0)^2/(2z_j)\right] $ where $k$
is the wavenumber of the beam. Hence we obtain
\begin{eqnarray}
\left\langle E_r^{*}(x)E_o(x)\right\rangle &=&\frac{\alpha _r^{*}\alpha
_oI_sk}{2\pi \sqrt{z_oz_r}}\exp [ik(z_o-z_r)]\int T(x_0)\exp \left[ \frac{ik%
}{2Z}(x-x_0)^2\right] dx_0  \label{4} \\
&\approx &(\alpha _r^{*}\alpha _oI_sk/\sqrt{2\pi z_oz_r})\exp
[ik(z_o-z_r)]\exp [ikx^2/(2Z)]\widetilde{T}(kx/Z).  \nonumber
\end{eqnarray}
Equation (\ref{4}) presents the Fresnel diffraction integral of an object
under the paraxial condition, the same as a coherent source does but with an
effective object distance $Z=z_oz_r/(z_r-z_o)$ replacing the real one $z_o$.
The approximation in Eq. (\ref{4}) is hold when the size of object is much
less than the area of diffraction pattern, and the Fourier transform $%
\widetilde{T}$ of object $T$ can be deduced, for instance, $\widetilde{T}%
(q)=(2b/\sqrt{2\pi })\sin $c$(qb/2)\cos (qd/2)$ for the double-slit.

In the above interference scheme, we have released the requirement of
spatial coherence, but still demand a better temporal coherence for the
source. This restriction can be relieved in the scheme of Fig. 1(b) in which
the two arms of the interferometer have the same distance while a lens of
the focal length $f_o$ is set at the middle position of the object path of
distance $2f_o$. In this configuration we obtain the interference term
\begin{eqnarray}
\left\langle E_r^{*}(x)E_o(x)\right\rangle &=&\frac{\alpha _r^{*}\alpha
_oI_sk}{2\sqrt{2}\pi f_o}\int T(x_0)\exp \left[ -\frac{ik}{4f_o}%
(x+x_0)^2\right] dx_0  \label{7} \\
&\approx &[\alpha _r^{*}\alpha _oI_sk/(2\sqrt{\pi }f_o)]\exp [-ikx^2/(4f_o)]%
\widetilde{T}[kx/(2f_o)],  \nonumber
\end{eqnarray}
which is equivalent to Eq. (\ref{4}). The experimental results of the
present scheme with $f_o=19cm$ are shown in Fig. 3, where (a) and (b)
exhibit the average intensity patterns $\langle I_1(x)\rangle $ and $\langle
I_2(x)\rangle $ registered by CCD$_1$ and CCD$_2$, respectively. We can see
that the two interference patterns having a phase shift $\pi $ are formed in
the sum of 10,000 frames and match with the theoretical simulation of Eq. (%
\ref{7}) in addition to an intensity background. Moreover, for a 50/50
beamsplitter BS$_2$, the difference and sum of the two patterns present the
net interference pattern and the intensity background, as shown in Figs.
3(c) and 3(d), respectively. As a matter of fact, the homogeneous intensity
background in Fig. 3(d) verifies the incoherence of the source. To further
confirm whether the interference pattern is related to the spatial
incoherence, we may compare it with the result obtained in the same
interferometer using coherent light. We simply remove the ground glass in
Fig. 1(b). In this case, the interference pattern for the coherent field
consists of two parts, $|\widetilde{T}(kx/f_o)|^2$ and $\widetilde{T}%
(kx/f_o)+$c.c.. The corresponding experimental results are plotted in Fig.
4, where (a) and (b) show the stable intensity patterns $I_1(x)$ and $I_2(x)$
registered by CCD$_1$ and CCD$_2$, respectively. After eliminating the
intensity of each arm, the net interference pattern in Fig. 4(c) fits the
formula $\widetilde{T}(kx/f_o)$, which has a doubled spatial frequency with
respect to that in Eq. (\ref{7}) for the incoherent source. Therefore, in
the same interferometer, both the coherent and incoherent sources can
perform Fourier transform of an object with different spatial frequency.

To further exploit the effect, we must consider a true thermal light source.
An extended thermal light source can be regarded as spatially incoherent
source with a short coherent time less than 0.1 nsec. Within the coherent
time, the source may produce an instantaneous exposure of interference
pattern in our schemes. Unlike the pseudothermal light source, each
individual exposure cannot be registered directly by the slow CCD camera
with the response time of order msec. Instead, an average intensity
distribution of these exposures will appear on the CCD\ screen. We have
indicated that the scheme of Fig. 1(b) is appropriate for observing
interference using ture thermal light source, since both the object and
reference waves travel the same distance and it thus releases the
requirement of temporal coherence. We use a Na lamp of wavelength 589.3$nm$
with the illumination area $10\times 10$ $mm^2$ to replace the pseudothermal
light source in Fig. 1(b) and find that the interference patterns directly
appear on the CCD screen, as shown in Fig. 5(b). For comparison, Fig. 5(a)
shows the 2D interference pattern corresponding to Fig. 3(a) for the
pseudothermal light in the same interferometer. The two fringes are similar
but with a slight different spacing, which displays different wavelength of
the two sources. Then we set a pinhole of diameter 0.36$mm$ after the lamp,
and the spatial incoherence has been dispelled. With this point-like source,
a different interference pattern, which has a half fringe spacing of that
for the spatially incoherent source, is recorded on the CCD screen, as shown
in Fig. 5(c).

We have both theoretically and experimentally demonstrated that a spatially
incoherent light source is capable of performing interference in an
unbalanced interferometer under certain configurations. Physically, each
point in the spatially incoherent source may produce an interference pattern
in the interferometer. A frame of sample pattern observed on the screen is
the incoherent superposition of those patterns corresponding to all
illuminating points in the incoherent source, and thus fluctuates randomly
due to the spatial incoherence. In most interferometric schemes so far, the
statistical average of the sample patterns will present a homogeneous
distribution. Our experimental results clarify that this obstacle can be
surpassed under certain interferometric configurations. Unexpectedly, in the
same interferometer the interference pattern for the spatially incoherent
source is well defined and equivalent to that for the coherent source but
with different spatial frequency. We note that, in the light of holography,
our approach is in essence different from the previous method called
''incoherent holography''\cite{my,lm,cr} which aims at encoding an
incoherent object, such as a fluorescence object. In the incoherent
holography, each source point in the object produces, by interfering its
wave fronts, a stationary two-dimensional intensity pattern (e.g. Fresnel
zone plate) which uniquely encodes the position and intensity of the object
point, and hence the method is limited to record intensity distribution for
the fluorescence object. However, in our approach, the hologram formed in
the statistical average of the patterns can be equivalent to that in the
coherent holography, recording the complete information of the object. The
present experiment can significantly refresh our intuition and experience:
the irregular phase distribution of incoherent field does not always wash
out the interference pattern. It is also interesting that photons emitted
from uncorrelated portions of the source can be cooperatively involved in a
well-defined interference pattern without photon spatial correlation. After
releasing the spatial coherence requirement, we may expect a wide and
potential application in the interference techniques especially for those
sources the coherence is unavailable, such as x-ray and electron sources.

This work was supported by the National Fundamental Research Program of
China, Project No. 2006CB921404, and the National Natural Science Foundation
of China, Project No. 10574015.


Figure Captions

Fig. 1. Experimental schemes of unbalanced interferometer using an
incoherent light source. P$_1$ and P$_2$ are two polarizers for modulating
intensity; G is a rotating ground glass; CCD$_1$ and CCD$_2$, two CCD
cameras. Two mirrors, M$_1$ and M$_2$, and two beamsplitters, BS$_1$ and BS$%
_2$, form an interferometer. T is a double-slit close to BS$_1$. (a) Two
arms have different distances; (b) One lens $L_o$ with the focal length $f_o$
is set at the middle position of the object arm, and two arms have the equal
distance $2f_o$.

Fig. 2. Experimental results of 2D interference patterns recorded by CCD$_1$
in the scheme of Fig. 1(a). (a) and (b) are individual single frames; (c),
(d), (e), (f) and (g) are averaged over 10, 40, 400, 6400 and 10,000 frames,
respectively.

Fig. 3. Experimental results of 1D interference patterns in the scheme of
Fig. 1(b). (a) and (b) are interference patterns (averaged over 10,000
frames) registered by CCD$_1$ and CCD$_2$, respectively; (c) and (d) are
their difference and summation, respectively. Experimental data and
theoretical simulation are given by open circles and solid lines,
respectively.

Fig. 4. Same as in Fig. 3 but removing the ground glass in Fig. 1(b). All
the interference patterns are stable.

Fig. 5. 2D interference patterns registered by CCD$_1$ in the scheme of Fig.
1(b). (a) with the original pseudothermal light source in Fig. 1(b); (b)
with a Na lamp of extended illumination area replacing the pseudothermal
light source; (c) with a Na lamp followed by a pinhole replacing the
pseudothermal light source.

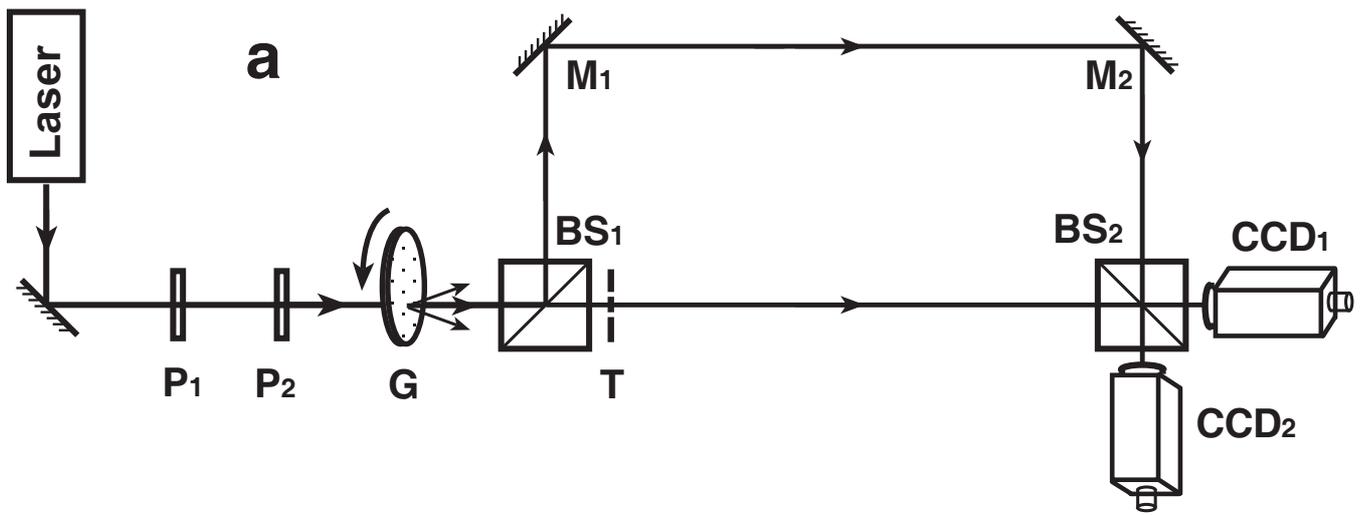
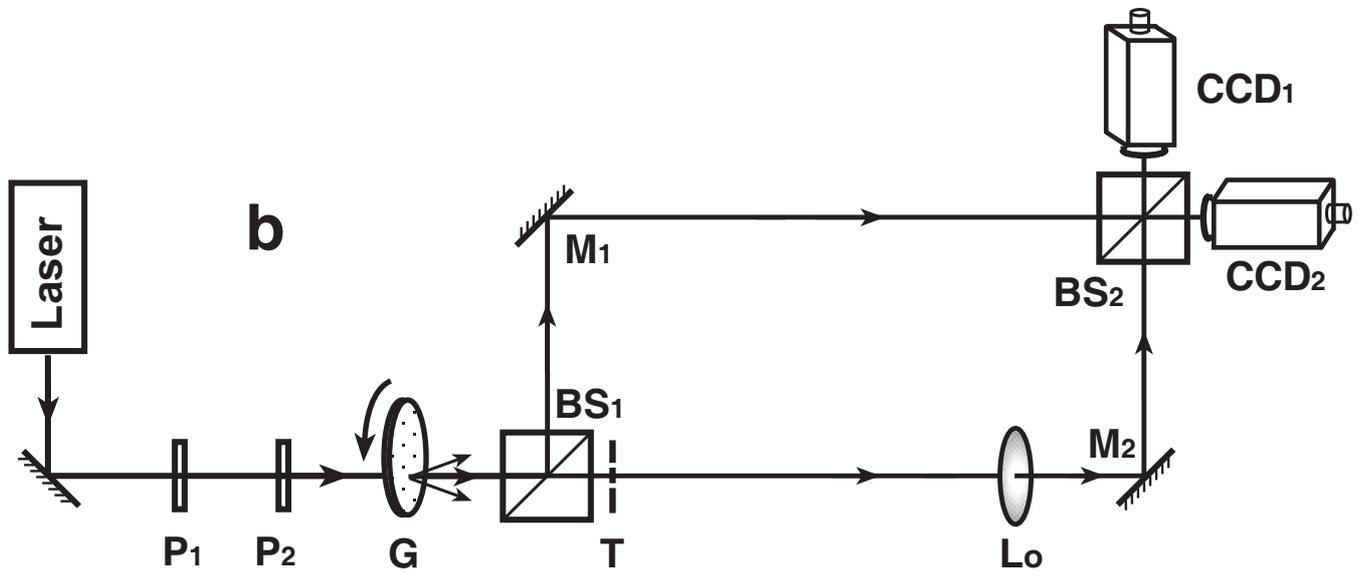
\end{document}